\documentstyle[aps,preprint,epsfig]{revtex}
\begin{document}
\draft
\title{Lattice Boltzmann model with hierarchical interactions}
\author{A. Lamura$^{1,a}$ and S. Succi$^{2}$}
\address{
$^1$ Istituto Applicazioni Calcolo, CNR, Sezione di Bari\\
Via Amendola 122/I, 70126 Bari, Italy \\
$^2$ Istituto Applicazioni Calcolo, CNR\\
V.le del Policlinico 137, 00161 Roma, Italy
}
\maketitle
\begin{abstract}
We present a numerical study of the dynamics of a non-ideal
fluid subject to a density-dependent pseudo-potential 
characterized by a hierarchy of nested attractive and repulsive interactions.
It is shown that above a critical threshold of the interaction
strength, the competition between stable and unstable regions 
results in a short-ranged disordered fluid pattern 
with sharp density contrasts.
These disordered configurations contrast with phase-separation
scenarios typically observed in binary fluids. 
The present results indicate that frustration can be modelled 
within the framework of a suitable one-body effective Boltzmann equation.
The lattice implementation of such an effective Boltzmann equation
may be seen as a preliminary step towards the development of 
complementary/alternative approaches to truly atomistic methods
for the computational study of glassy dynamics.
\end{abstract}
\noindent 
$^a$ Corresponding author
E\_mail: a.lamura@area.ba.cnr.it Phone: +39 0805530719 Fax: +39 0805588235

\section{Introduction}

A deeper understanding of the behavior of 
disordered systems, and notably glassy ones,
represents an outstanding challenge
in modern condensed matter physics.
In the face of the lack of a comprehensive theory, much work is currently
devoted to the numerical simulation of glassy behavior.
To date, these simulation efforts rely mainly on Molecular
Dynamics and Monte Carlo techniques \cite{GLAP,GLAR}.
Both methods incorporate details of microscopic behavior
(intermolecular potentials) and, as a result, they fall short of
reaching spatio-temporal scales of macroscopic interest.
A typical molecular dynamics simulation would last a few 
tens-hundreds of nanoseconds, to be contrasted 
with relaxation times of the order of an hour and more of real glasses.
Larger scales are made accessible by coarse-grained lattice gas models
in which the details of true molecular interactions are replaced
by suitable heuristics about the behavior of fictitious lattice
molecules. These models reach out larger scales but the 
corresponding outcomes have to be critically weighted
against the simplifying assumptions they are based upon.
At the extreme of this line one finds spin-glasses 
\cite{GLA,SG1} and their modern lattice glass variants \cite{SG2,LGL}.

Although to a different degree of physical fidelity, all of these models
retain the many-body nature of the intermolecular interactions, with
an inevitable price in terms of computational effort.
It is therefore of great interest to explore whether mesoscopic models,
namely suitable {\it effective one-body formulations} may capture at least
some basic features of glassy behavior, at a fraction of the computational
cost associated with the aforementioned many-body simulation techniques.

In the present paper this task is pursued within the framework of the
lattice Boltzmann equation (LBE), a minimal
kinetic equation describing stylized pseudo-molecules 
evolving in a regular lattice according to a simple 
local dynamics including free-streaming, collisional relaxation 
and (effective) intermolecular interactions \cite{LBE,PROCS,MIL}.
In particular, we shall show that it is
possible to construct a class of suitable mesoscopic
potentials supporting some basic traits of disordered systems.
Here we shall focus on a minimal requirement: 
The ability to sustain sharp density contrasts
within a disordered pattern.

The paper is organized as follows. 
In the next Section, after a cursory view of the main features 
of the standard LBE, we present our model.
Section III is devoted to the results of numerical 
simulations and finally, we discuss the main findings 
of the present paper and outline possible future 
research directions.

\section{The model}

The most popular form of lattice Boltzmann equation
(Lattice BGK, for Bhatnagar, Gross, Krook)
\cite{LBGK}, reads as follows:
\begin{equation} 
f_i(\vec{r}+\vec{c}_i,t+1)-f_i(\vec{r},t)= -\omega[f_i-f_i^e](\vec{r},t)
\end{equation}
where $f_i(\vec{r},t) \equiv f(\vec{r},\vec{v}=\vec{c}_i,t)$ is a
discrete distribution function of particles
moving along with discrete speed $\vec{c}_i$.
The right-hand side represents
the relaxation to a local equilibrium $f_i^e$
in a time lapse of the order of $\omega^{-1}$.
This local equilibrium is usually taken in the form
of a quadratic expansion of a local Maxwellian:
\begin{equation}
f_i^e = \rho w_i [1+\frac{\vec{u} \cdot \vec{c}_i}{c_s^2}
+\frac{\vec{u}\vec{u} \cdot (\vec{c}_i \vec{c}_i-c_s^2 I)}{2 c_s^4}]
\end{equation}
where $\rho=\sum_i f_i$ is the fluid density,
$\rho \vec{u} = \sum_i f_i \vec{c}_i$ the fluid
current, $w_i$ is a set of weights normalized to unity,
$c_s$ the lattice sound speed (a constant equal to $1/\sqrt 3$ in
our case) and finally, $I$ stands for the unit tensor.
The set of discrete speeds must be chosen in such a way as to guarantee
mass, momentum and energy conservation, as well as rotational invariance.
Only a limited subclass of lattices qualifies.
In the sequel, we shall refer to the nine-speed lattice consisting
of zero-speed, speed one $c=1$ (nearest-neighbor connection), and
speed $c=\sqrt 2$, (next-nearest-neighbor connection).
Standard analysis shows that the LBE system behaves like a 
quasi-incompressible fluid with an ideal equation
of state $p=\rho c_s^2$ and kinematic viscosity
$\nu = c_s^2(1/\omega-1/2)$.
Departures from ideal-gas behavior within the LBE
formalism are typically described by means of phenomenological
pseudo-potentials mimicking the effects of potential energy
interactions \cite{SC}. 
These pseudo-potentials have proved capable of describing a wide
range of complex fluid behaviors, such as phase-separation
in binary fluids and phase-transitions. 
To date, however, no LBE model appears 
to have addressed frustrated systems. 
This is precisely the scope of this work.

We choose the empirical pseudo-potential in the form of
a lattice-truncated repulsive Lennard-Jones
interaction:
\begin{eqnarray}
V(r) & = & G \Psi(\rho) (r/\sigma)^{-12},\;\;\; 1<r<\sqrt 2\nonumber \\
V(r) & = & 0,\;\;\;\;\;\;\;\;\;\;\;\;\;\;\;\;\;\;\;\;\;\;\;\;\;\;\;\; elsewhere
\end{eqnarray}
where the cut-off $\sigma=1$ in lattice units
and $G$ is a parameter controlling the strength
of the interaction (an inverse effective temperature), hence
the surface tension between dense and light phases.
The density-dependence is taken in the form
of a hierarchy of polynomials, whose density zeroes are distributed
according to a binary tree of depth $N_g$ (number of generations):
\begin{equation}
\Psi(\rho) =(\rho-\rho_0)
\; \prod_{g=2}^{N_g} 
\frac{-1}{(1+(\frac{\rho-\rho_0}{\delta \rho_0/2})^{2^{g-1}})^2}
\prod_{k=1}^{2^{g-1}} 
\frac{(\rho-\rho_{gk})}{(\rho_{gk}-\rho_0)}
\end{equation}
where $\delta \rho_0$ is the typical scale for the density gap.
Starting with the root value $\rho_0$, at
each generation $g$, a mother density $\rho_{gk}$
spawns two children $\rho_{g+1,2k},\rho_{g+1,2k+1}$
spaced by an amount $\pm \delta \rho_0/2^{g-1}$ 
away from the mother value.
This hierarchical potential implements 
the presence of $N_{\rho}=2^{N_g}-1$ competing density extrema.

As an example, for $N_g=3$,
\begin{eqnarray}
g=1 &:& \rho_0 \nonumber \\
g=2 &:& \rho_{1,2}=\rho_0 (1 \mp \delta \rho_0/2) \\
g=3 &:& \rho_{3,4,5,6}=\rho_0 (1 \mp \delta \rho_0/2 \mp \delta \rho_0/4) \nonumber 
\end{eqnarray}
The denominator serves the purpose of letting $\Psi \rightarrow 0$
outside the hierarchical range of zeroes $\rho_{gk}$.
The result is an effective potential alternating stable repulsive regions 
($\Psi'>0$, prime meaning derivative with respect to $\rho$) 
with unstable attractive ones ($\Psi'<0$),
distributed approximately in the density range
$\rho_{\mp}=\rho_0 (1 \mp \delta \rho_0 (1-2^{-N_g}))$.
Since the case $N_g=2$ (cubic interaction) is somewhat equivalent
to a Van der Waals interaction, we can think to our
potential as of a series 
of hierarchically nested van der Waals interactions. 
It is intuitively clear that if the coupling strength
$G$ is sufficiently strong, attractive regions are
subject to instabilities which can be likened to a phase transition.
However, since there are multiple forbidden regions 
interweaved with density gaps with (meta)stable behavior, 
the system is amenable to a full cascade of phase-transitions,
depending on the strength of the interaction.
As a result, the fluid is expected to exhibit a
sort of ``frustration'' resulting from the competition
of the multiple density minima.
This competition is resolved through steep interfaces
connecting the stable density regimes.
In the language of free-volume theory, this frustration
relates to the disordered short-range coexistence of
two competing species: ``voids'' (regions with $\rho<\rho_0$)
and ``cages'' (regions with $\rho>\rho_0$).
This contrasts with binary separation scenarios, in which the two
species organize over coherent patterns (``blobs'')
of sizeable extent.

The LBE with hierarchical interactions takes the final form
\begin{equation}
\label{GLBE}
f_i(\vec{r}+\vec{c}_i,t+1)-f_i(\vec{r},t)= -\omega[f_i-f^{eq}_i](\vec{r},t)
+ F_i[\rho(\vec{r},t)]
\end{equation}
where $F_i = - G \frac{\rho}{\rho_0} \nabla \Psi [\rho] \cdot \vec{c}_i$.
This can be regarded as a dynamic, mean-field model of 
a non-ideal fluid with multiple density phases.

\section{Numerical simulations}

\begin{figure}[h]
\begin{center}
\includegraphics[width=9.2cm,angle=0]{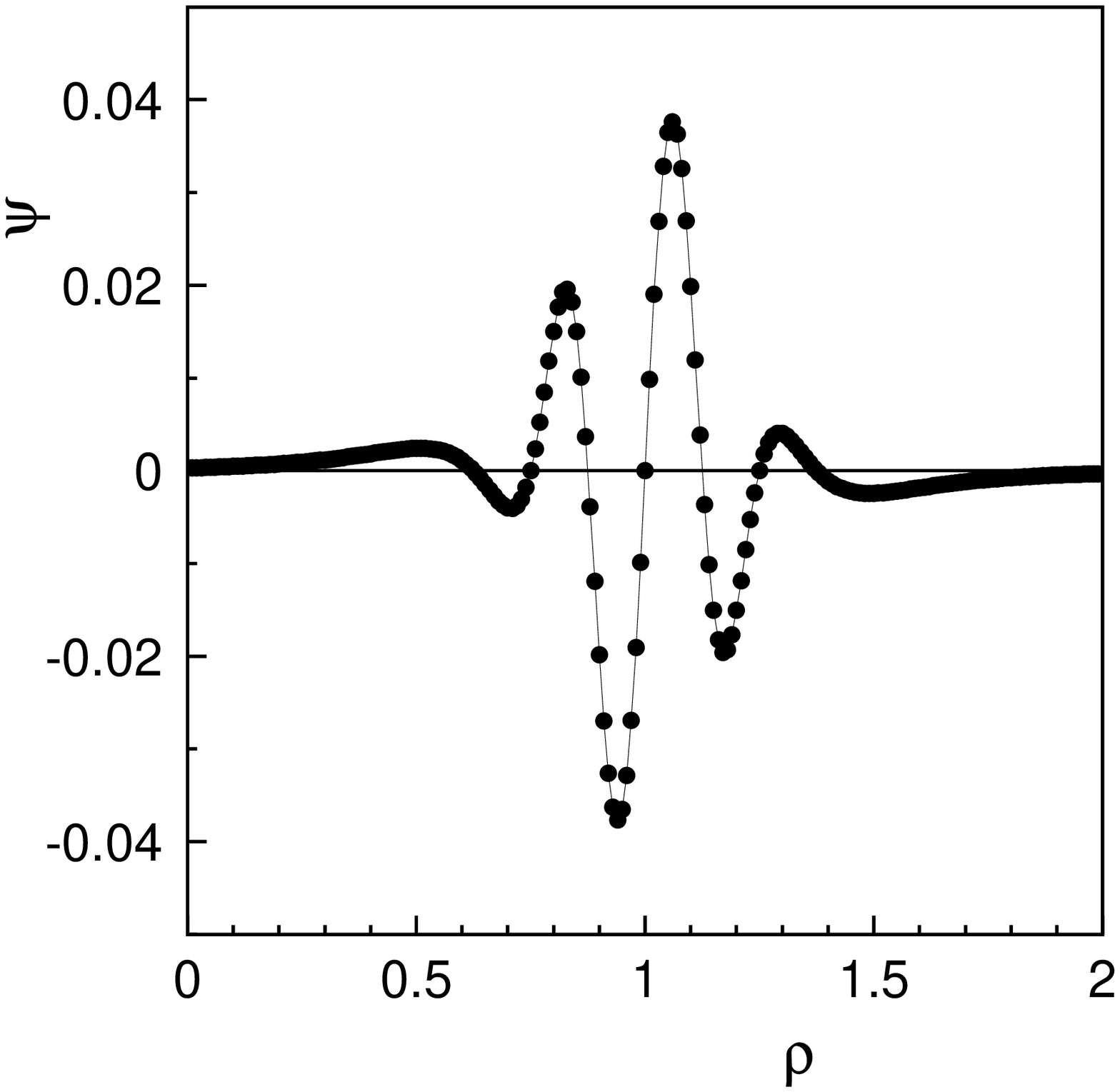}
\end{center}
\caption{The pseudo-potential $\Psi(\rho)$ for the case
$N_g=3$, $\rho_0=1.0$, $\delta \rho_0=0.5$.}
\label{FIG1}
\end{figure}

\begin{figure}[h]
\begin{center}
\includegraphics[width=9.2cm,angle=0]{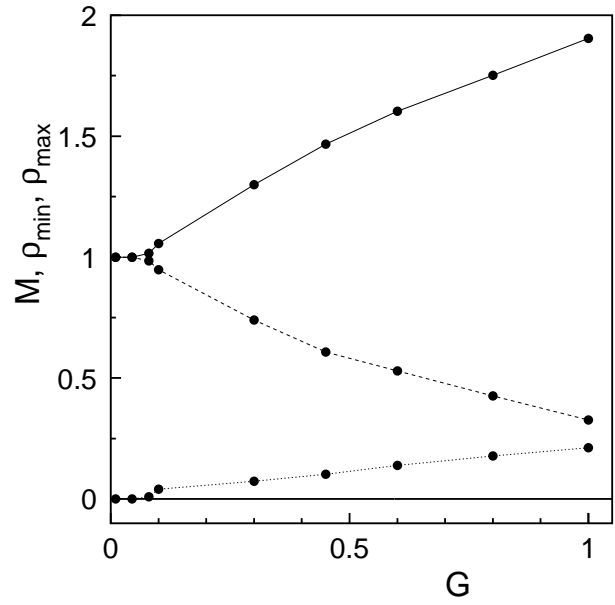}
\end{center}
\caption{The maximum (full line) and minimum (dashed line) densities
and the order parameter $M$ (dotted line) as a function
of the coupling strength $G$.}
\label{FIG2}
\end{figure}

\begin{figure}
\begin{center}
\includegraphics[width=9.2cm,angle=0]{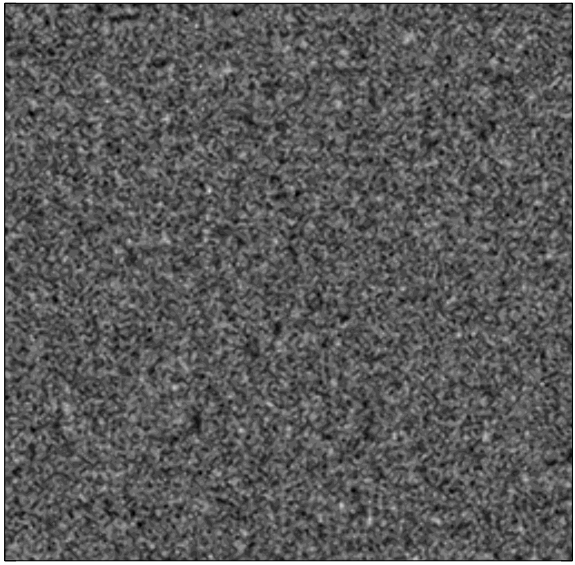}
\end{center}
\caption{Snapshot of the density configuration at $t=200$
for the case $G=0.45$. Black/white colors code for high/low density.}
\label{FIG3}
\end{figure}

To assess the viability of the present LBE, we have simulated
a 2D dimensional fluid with the following parameters:
Grid size $256 \times 256$, $\omega=1.0$, 
$\rho_0 =1.0$, $N_g=3$, $\delta\rho_0 =0.5$, and
$G$ is varied between $0$ and $1.0$.
The resulting effective potential $\Psi(\rho)$
is shown in Fig.~\ref{FIG1}.
The initial condition is $\rho(x,y)=\rho_0 \; (1+\xi)$
where $\xi$ is a random perturbation uniformly
distributed in the range $[-0.01:0.01]$.
Due to the combined effect of kinetic hopping 
from site to site (the left-hand-side
of the LBE) and mode-mode coupling \cite{GOTZE} induced by the
non-linear potential $\Psi(\rho)$, the density distribution
spreads out in time, so that an instability
is triggered as soon as $G$ exceeds the critical threshold $G_c$.
In Fig.~\ref{FIG2} we show the maximum and minimum densities
as a function of the coupling strength $G$.
This Figure highlights a sizeable symmetry breaking 
of the order of $\Delta \rho/\rho >1$ for $G>G_c$, where
$\Delta \rho = \rho_{max}-\rho_{min}$.
The density gap $\Delta \rho$ starts up at $G=G_c \sim 0.09$ and
rapidly fills the available range of states between
$\rho_{-}$ and $\rho_{+}$, as the interaction strength is increased.
The scaling law is approximately $\Delta \rho \sim (G-G_c)^{1/2}$ 
in accordance with mean-field theory. 
The critical value estimated on the basis of the stability condition
\begin{equation}
|\Psi'| > c_s^2/G
\end{equation}
yields $G_c \sim 0.089$, in a nice agreement with numerical data.
A similar message is sent by the order parameter
$M=\sqrt{<(\delta \rho)^2>}/<\rho>$, where 
$\delta \rho \equiv \rho-<\rho>$
is the density fluctuation around its spatial average,
which is
represented by the dotted line in Fig.~\ref{FIG2}.
Inspection of the spatial distribution of the density
field $\rho(x,y)$ shows that this non-zero order parameter
is not associated with phase-separation, but
rather with a disordered coexistence of voids 
and cages (see Fig.~\ref{FIG3}).
This shows that the potential allows
to model sharp 
density contrasts over a disordered spatial distribution.

\begin{figure}
\begin{center}
\includegraphics[width=9.2cm,angle=0]{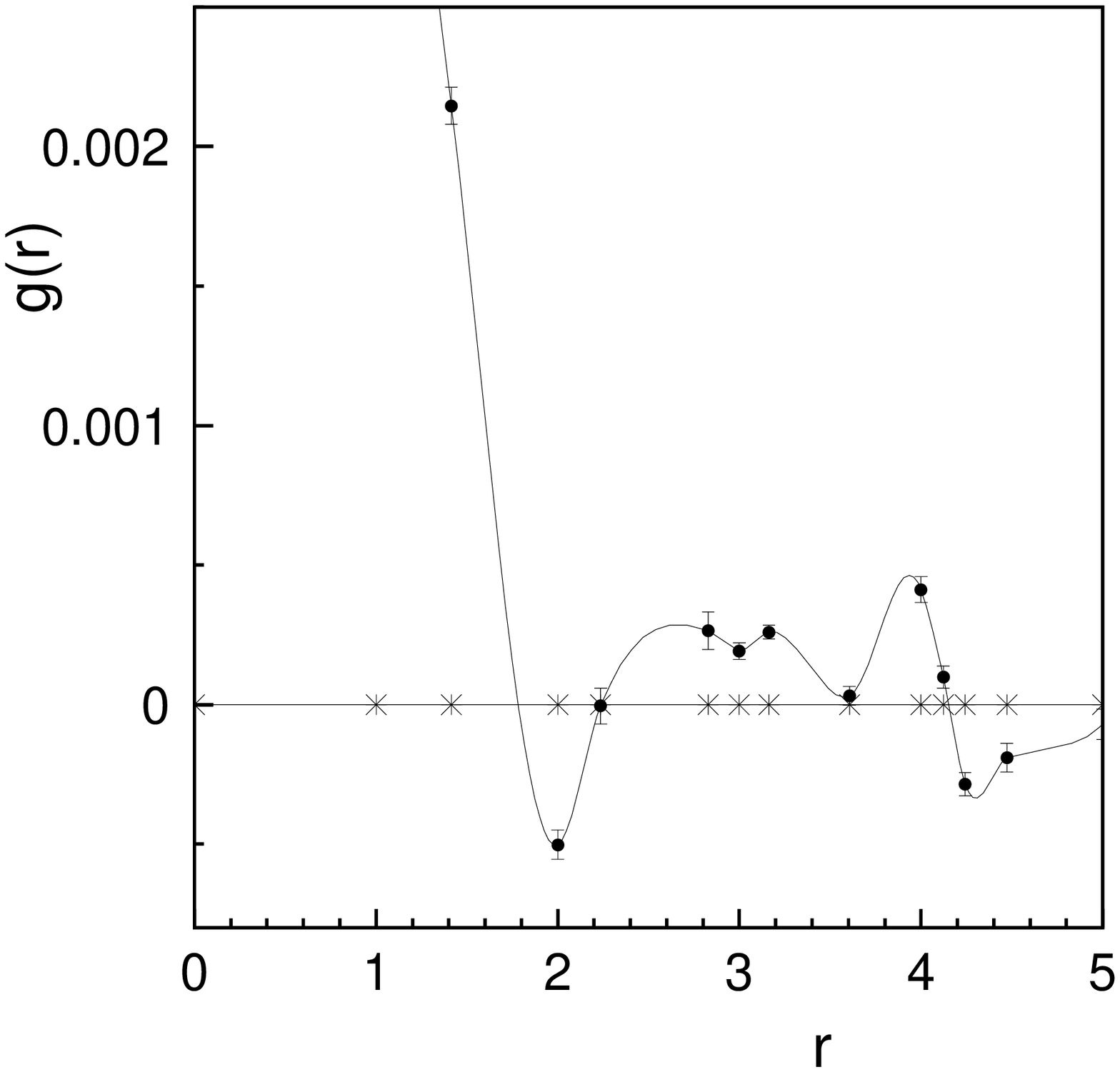}
\end{center}
\caption{Density-density correlation function at $t=200$
for the cases $G=0$ ($\ast$) and $G=0.45$ ($\bullet$), averaged over ten
independent runs. Error bars are shown for the case $G=0.45$.}
\label{FIG4}
\end{figure}

As a further observable, in Fig.~\ref{FIG4} we plot the 
density-density correlation:
\begin{equation}
g(r)=\frac{<\delta \rho(x+r_x,y+r_y) \delta \rho(x,y)>}
{<\rho(x,y) \rho(x,y)>},
\end{equation}
where $r^2=r_x^2 + r_y^2$.
The curve has been obtained upon averaging the quantities $g(r)$
over ten independent runs with
different initial configurations.
The error bars show the maximum errors on the values $g(r)$ at each value
of $r$.
This function exhibits oscillating-decay behavior.
Being in a lattice, it is clear that data are only available at discrete
positions $1,\sqrt 2, 2, \dots$. 
The small error bars and the non-oscillating
behavior of $g(r)$ in the absence of potential,
which is virtually zero as a consequence of the uniform and
constant density distribution, suggest that the observed oscillations in the
case $G \neq 0$ are 
not a numerical artifact. However, only further simulations of larger 
systems can confirm this result.

\begin{figure}
\begin{center}
\includegraphics[width=9.2cm,angle=0]{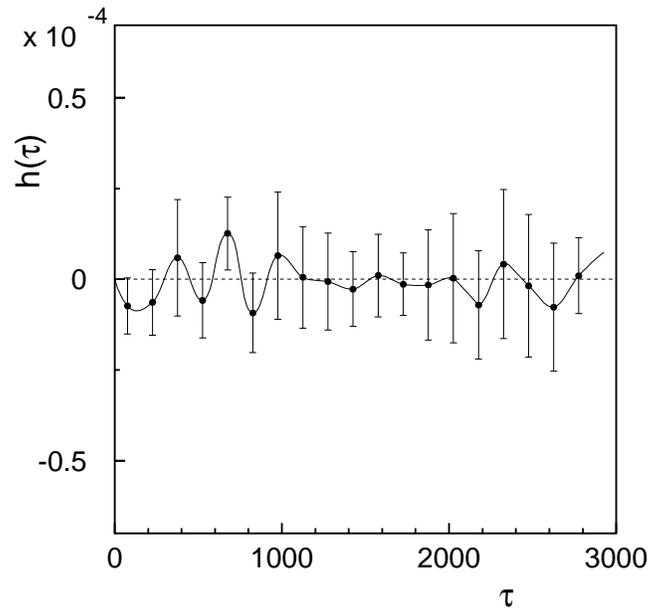}
\end{center}
\caption{Density-density time correlation function
for the cases $G=0$ (dashed line) and $G=0.45$ (full line), averaged over ten
independent runs. Error bars are shown for the case $G=0.45$.}
\label{FIG5}
\end{figure}

In order to look for the existence of 
slow relaxation modes, we computed the density-density
time-correlation function $h(\tau)$ (see Fig.~\ref{FIG5})
\begin{equation}
h(\tau)=\frac{<\delta \rho(x,y,t+\tau) \delta \rho(x,y,t)>_t}
{<\rho(x,y,t) \rho(x,y,t)>_t}
\end{equation}
taken at $x=128,y=128$. In the above, $<...>_t$ denotes a time average.
We have run ten independent simulations starting from different initial 
configurations, and we computed $h(\tau)$ in each run. 
These individual quantities were first averaged
over time intervals of $150$ time-steps in order to eliminate high frequencies,
and then the final average over the ten runs was taken. 
These quantities are plotted
in Fig.~\ref{FIG5} where the error bars represent the maximum error on
the values of $h(\tau)$. 
We find persistent oscillations around zero, possibly reflecting an everlasting
competition between the various density minima of
the potential over the timescale of the simulation.
The case $G=0$ shows an immediate decay to zero of the time-correlation
function, reflecting the fast relaxation towards a uniform distribution
when no potential is acting on the system.

\section{Conclusions}

We have presented a new mesoscopic lattice Boltzmann 
model with a density-dependent hierarchy of attractive and
repulsive interactions. 
The long-time dynamics of this model yields disordered patterns 
with sharp short-range density contrasts, as opposed to binary 
phase-separation scenarios. 
Such configurations do not relax towards a
frozen state, but continuously change  
preserving a disordered spatial structure.
The present results show that one distinguishing feature 
of disordered behavior, namely frustration, can be modeled 
within the framework of a suitable one-body effective kinetic theory.
On the other hand, the same results also show that further crucial signatures
of glassy behavior, such as slow time-relaxation, are not captured
by the present hierarchical model.
At this point, the main question is: what major ingredients of glassy
dynamics are we missing in the present model?

According to Palmer et al \cite{PALM}, a successfully theory
of glassy relaxation should satisfy three basic requirements:
(1) The theory must be based on {\it dynamics}, and not just statistics,
because since glasses break ergodicity, equilibrium distributions are
of scanty use.
(2) The above dynamics must be {\it constrained}, because there is 
no reasonable
hope to diagonalize such highly non-linear systems into independent modes.
And finally, (3) the theory should be {\it hierarchical}, so that
slow modes can adiabatically enslave the fast ones, while the fast modes, in
return, set constraints on the dynamics of the slow ones.

Our hierarchical LB model does indeed 
{\it generically} meet with all of these
three requirements, but only incompletely.
The model is dynamic, but nonetheless based on a single-time
relaxation to local equilibrium.
Our model is also non-linear and hierarchically constrained, but still
within the class of effective one-body theories. 
Such effective one-body theories prove exceedingly successful for
simple fluids, but it might be that the physics of glasses, and particularly
dynamic inhomogeneity \cite{HET1,HET2}, is just too intimately 
related to many-body effects which are beyond the reach 
of effective mean field theories, no matter how ingenious.
We believe that, independently of the computational pay-offs discussed
earlier in this work, this latter question bears a significant
theoretical interest on its own.
Work in this direction is currently underway.

{\bf Acknowledgments}\\
We thank Professors E. Marinari and G. Parisi 
for many illuminating comments and discussions.
SS wishes to thank Prof. K. Binder for valuable discussions and
for bringing Ref.~\cite{GLAP} to his attention.

\newpage

\newpage

\newpage

\newpage

\newpage

\end{document}